\documentclass[fleqn,10pt]{SelfArx} % Document font size and equations flushed left

\usepackage[english]{babel} % Specify a different language here - english by default

\usepackage{lipsum} % Required to insert dummy text. To be removed otherwise

\definecolor{color1}{RGB}{0,0,90} % Color of the article title and sections
\definecolor{color2}{RGB}{0,20,20} % Color of the boxes behind the abstract and headings

\usepackage{hyperref} % Required for hyperlinks
\hypersetup{hidelinks,colorlinks,breaklinks=true,urlcolor=color2,citecolor=color1,linkcolor=color1,bookmarksopen=false,pdftitle={Title},pdfauthor={Author}}

\usepackage{adjustbox}

\usepackage{changepage}

\JournalInfo{~} % Journal information
\Archive{~} % Additional notes (e.g. copyright, DOI, review/research article)

\PaperTitle{A Short Introduction to Information-Theoretic Cost-Benefit Analysis} % Article title

\Authors{Min Chen}
\affiliation{\textit{University of Oxford, United Kingdom}}
\affiliation{*\textbf{Corresponding author}: min.chen@oerc.ox.ac.uk}
 % Keywords - if you don't want any simply remove all the text between the curly brackets
\newcommand{\keywordname}{Keywords} % Defines the keywords heading name

\Abstract{This arXiv report provides a short introduction to the information-theoretic measure proposed by Chen and Golan in 2016 for analyzing machine- and human-centric processes in data intelligence workflows. This introduction was compiled based on several appendices written to accompany a few research papers on topics of data visualization and visual analytics. Although the original 2016 paper and the follow-on papers were mostly published in the field of visualization and visual analytics, the cost-benefit measure can help explain the informative trade-off in a wide range of data intelligence phenomena including machine learning, human cognition, language development, and so on. Meanwhile, there is an ongoing effort to improve its mathematical properties in order to make it more intuitive and usable in practical applications as a measurement tool.}

\begin{document}

\flushbottom % Makes all text pages the same height

\maketitle % Print the title and abstract box

\tableofcontents % Print the contents section

\thispagestyle{empty} % Removes page numbering from the first page

%----------------------------------------------------------------------------------------
%	ARTICLE CONTENTS
%----------------------------------------------------------------------------------------

% ====================
\section{Introduction}
In the autumn of 2014, the director of Info-Metrics Institute, Professor Amos Golan (American University, DC) invited me to visit him for a few weeks. We worked on a few blue sky ideas in information-theoretic research. One of them was proposed by Amos: ``\emph{Stock market data is now in microseconds, and may soon be in nanoseconds. As an economist, I do not need to work at such a data resolution. What is the optimal resolution for an economist?}'' We met every weekday in Amos's office and brainstormed various concepts and measurement for informative optimization.

During my visit, I realized that a visualization process might have something in common with a statistical aggregation function or a more complex statistical inference process. If we could define any statistical process or visualization process as a transformation from an input information space to an output information space, Amos's question would also apply to visualization.

During the Fall 2014 Conference of the Info-Metrics Institute, I mentioned to Amos that there might be a trade-off formula for answering the question. Amos was very busy in organizing the conference, and we did not manage to discuss it further until early 2015 when we started to work on a submission to IEEE SciVis. Amos suggested to name the formula as a \emph{cost-benefit ratio}.

The submission was not accepted by SciVis 2015. Huamin Qu (HKUST) and Chris Johnson (Utah) encouraged me to submit it to IEEE TVCG. It received a decision of a major revision, including a request to conduct an empirical study to prove that the cost-benefit ratio is correct. Based on my experience of conducting empirical studies, I knew that this would not be a trivial undertaking. We submitted a revision without the required empirical study, though we did follow the request of some reviewers to add a new section on how to falsify the proposed formula. Luckily, the reviewers were open-minded and understanding, and accepted the paper \cite{Chen:2016:TVCG} without insisting on an empirical study.

Although visualization designers make trade-off decisions all the time, the abstraction of such judgement as a formula has not been easy to understand. When several pieces of follow-on work were submitted for review, reviewers often asked for more explanation of the cost-benefit ratio. I have written appendices to accompany a number of submissions (e.g., \cite{Chen:2019:CGF,Chen:2019:TVCG, Tennekes:2021:CGF,Chen:2021:arXiv:T,Chen:2021:arXiv:E}).
This introduction was compiled based on these appendices.
Hopefully, the text is easier to access than those appendices that may be difficult to notice at the publishers' web sites.

\paragraph{Acknowledgement.} I would like to thank Amos Golan for our joint work on the cost-benefit ratio \cite{Chen:2016:TVCG} and my co-authors of the follow-on papers who have proofread the relevant explanatory texts in the papers and appendices, including David Ebert \cite{Chen:2019:CGF},
Kelly Gaither and Nigel W. John \cite{Chen:2019:TVCG},
Martijn Tennekes \cite{Tennekes:2021:CGF},
Mateu Sbert \cite{Chen:2021:arXiv:T}, and Alfie Abdul-Rahman \cite{Chen:2021:arXiv:E}.
I am also grateful to the comments and revision suggestions made by anonymous reviewers.

% ====================
\section{\textbf{The Cost-Benefit Measure}}
\label{sec:OriginalTheory}
This section contains an extraction from a previous publication \cite{Chen:2019:CGF}, which provides a relatively concise but informative description of the cost-benefit ratio proposed in \cite{Chen:2016:TVCG}. The extraction has been modified slightly.

Chen and Golan introduced an information-theoretic measure for analyzing the cost-benefit ratio of a visual analytics (VA) workflow or any of its component processes \cite{Chen:2016:TVCG}.
The cost-benefit ratio consists of three fundamental measures that are abstract representations of a variety of qualitative and quantitative criteria used in practice, including
operational requirements (e.g., accuracy, speed, errors, uncertainty, provenance, automation),
analytical capability (e.g., filtering, clustering, classification, summarization),
cognitive capabilities (e.g., memorization, learning, context-awareness, confidence), and so on.
The abstraction results in a metric with the desirable mathematical simplicity \cite{Chen:2016:TVCG}.
The qualitative form of the metric is as follows:
\begin{equation}
\label{eq:CBR}
\frac{\textit{Benefit}}{\textit{Cost}} = \frac{\textit{Alphabet Compression} - \textit{Potential Distortion}}{\textit{Cost}}
\end{equation}

The measure describes the trade-off among the three fundamental measures: \emph{Alphabet Compression} (AC), \emph{Potential Distortion} (PD), and \emph{Cost} (Ct).

% --------------------
\subsection{Alphabet Compression (AC)}

\emph{Alphabet Compression} (AC) measures the amount of entropy reduction (or information loss) achieved by a process.
As it was noticed in \cite{Chen:2016:TVCG}, most visual analytics processes (e.g., statistical aggregation, sorting, clustering, visual mapping, and interaction) feature many-to-one mappings from input to output, hence losing information.
Although information loss is commonly regarded harmful, it cannot be all bad if it is a general trend of VA workflows.
Thus the cost-benefit ratio makes AC a positive component.

As soon as we measure the positive aspect of information loss, it becomes much easier to explain many data intelligence processes such as statistics, algorithms, visualization, and human decision-making are useful in principle. Such processes mostly feature many-to-one mappings, thus information loss. In the original paper \cite{Chen:2016:TVCG}, Chen and Golan illustrated such information loss using a simple workflow from receiving stock market data to deciding if one should buy, sell, or hold the shares of a particular stock. As the machine- and human-centric processes in the workflow all lose information rapidly, it would not make sense until considering the positive aspect of information loss.

% --------------------
\subsection{Potential Distortion (PD)}

\emph{Potential Distortion} (PD) balances the positive nature of AC by measuring the errors typically due to information loss. Instead of measuring mapping errors using some third party metrics or functions, PD measures the potential distortion when one reconstructs inputs from outputs.
The measurement takes into account humans' knowledge that can be used to improve the reconstruction processes. For example, given an average mark of 62\%, the teacher who taught the class can normally guess the distribution of the marks among the students better than an arbitrary person.

In many scenarios, there is no agreeable party metrics or functions about the errors in a process, i.e., in mapping from inputs to outputs. When we were working the cost-benefit ratio, Amos Golan, the co-author of \cite{Chen:2016:TVCG}, remarked: in economics, a decision might seem correct today, but may not be considered correct next month, next year, or next decade.  Hence measuring the divergence of the reconstruction is independent from any third party criteria.
The introduction of the notion of reconstruction also brings data intelligence processes into line with other informative processes in communication, compression, and encryption \cite{Chen:2020:OUP}.

% --------------------
\subsection{Cost (Ct)}

\emph{Cost} (Ct) of the forward transformation from input to output and the inverse transformation of reconstruction provides a further balancing factor in the cost-benefit metric in addition to the trade-off between AC and PD. The fundamental measurement of the cost is the amount of energy required to perform the actions of the process, including all activities for reconstructing inputs from outputs if such activities are present in the process. In practice, one may approximate the cost using \emph{time} or a monetary measurement.

% ====================
\section{It can Explain Why Visualization is Useful}
\label{sec:Why}
The cost-benefit measure was first published in the field of visualization \cite{Chen:2016:TVCG}.
One objective was to explain why visualization is useful in a mathematical way.
When the paper was reviewed, first by IEEE SciVis 2015 and then IEEE TVCG, some reviewers wanted a proof that this would be a correct measurement.
Most measurement systems are not ground truth.
They are functions that map some reality to some quantitative values, in order to aid the explanation of the reality and the computation of making predictions.
The cost-benefit measure proposed by Chen and Golan is one of such functions.
Following the request of some reviewers, the original paper contains a section about how to falsify the cost-benefit measure.

In this section, we provide a relatively informal and somehow conversational discussion about using this measure to explain why visualization is useful.

There have been many arguments about why visualization is useful. 
Streeb et al. collected a large number of arguments and found many arguments are in conflict with each other \cite{Streeb:2021:TVCG}.
Chen and Edwards presented an overview of schools of thought in the field of visualization, and showed that the ``why'' question is a bone of major contention \cite{Chen:2020:book}.

The most common answer to ``why'' question is because visualization offers insight or helps humans to gain insight. When this argument is used outside the visualization community, there are often counter-arguments that statistics and algorithms can offer insight automatically and often with better accuracy and efficiency. There are also concerns that visualization may mislead viewers, which cast further doubts about the usefulness of visualization, while leading to a related argument that ``visualization must be accurate'' in order for it to be useful.

The accuracy argument itself is not bullet-proof since there are many types of uncertainty in a visualization process, from uncertainty in data, to that caused by visual mapping, and to that during perception and cognition \cite{Dasgupta:2012:CGF}.
Nevertheless, it is easier to postulate that visualization must be accurate, as it seems to be counter-intuitive to condone the idea that ``visualization can be inaccurate,'' not mentioning the idea of ``visualization is normally inaccurate,'' or ``visualization should be inaccurate.''

The word ``inaccurate'' is itself an abstraction of many different types of inaccuracy.
Misrepresentation truth is a type of inaccuracy.
Such acts are mostly wrong, but some (such as wordplay and sarcasm) may cause less harm.
Converting a student's mark in the range of [0, 100] to the range of [A, B, C, D, E, F] is another type of inaccuracy.
This is a common practice, and must be useful.
From an information-theoretic perspective, these two types of inaccuracy are information loss.

In their paper \cite{Chen:2016:TVCG}, Chen and Golan observed that statistics and algorithms usually lose more information than visualization. Hence, this provides the first hint about the usefulness of visualization. They also noticed that like wordplay and sarcasm, the harm of information loss can be alleviated by knowledge. For someone who can understand a workplay (e.g., a pun) or can sense a sarcastic comment, the misrepresentation can be corrected by that person at the receiving end. This provides the second hint about the usefulness of visualization because any ``misrepresentation'' in visualization may be corrected by a viewer with appropriate knowledge.

On the other hand, statistics and algorithms are also useful, and sometimes more useful than visualization. Because statistics and algorithms usually cause more information loss, some aspects of information loss must be useful.
One important merit of losing information in one process is that the succeeding process has less information to handle, and thus incurs less cost.
This is why Chen and Golan divided information loss into two components, a positive component called \emph{alphabet compression} and a negative component called \emph{potential distortion} \cite{Chen:2016:TVCG}.

The positive component explains why statistics, algorithms, visualization, and interaction are useful because they all lose information.
The negative component explains why they are sometimes less useful because information loss may cause distortion during information reconstruction.
Both components are moderated by the cost of a process (i.e., statistics, algorithms, visualization, or interaction) in losing information and reconstructing the original information.
Hence, given a dataset, the best visualization is the one that loses most information while causing the least distortion.
This also explains why visual abstraction is effective when the viewers have adequate knowledge to reconstruct the lost information and may not be effective otherwise \cite{Viola:2019:book}.

The central thesis by Chen and Golan \cite{Chen:2016:TVCG} may appear to be counter-intuitive to many as it seems to suggest ``inaccuracy is a good thing'', partly because the word ``inaccuracy'' is an abstraction of many meanings and itself features information loss. Perhaps the reason for the conventional wisdom is that it is relatively easy to think that ``visualization must be accurate''. To a very small extent, this is a bit like the easiness to think ``the earth is flat'' a few centuries ago, because the evidence for supporting that wisdom was available everywhere, right in front of everyone at that time.
Once we step outside the field of visualization, we can see the phenomena of inaccuracy everywhere, in statistics and algorithms as well as in visualization and interaction.
All these suggest that ``the earth may not be flat,'' or ``inaccuracy can be a good thing.''

In summary, the cost-benefit measure by Chen and Golan \cite{Chen:2016:TVCG} explains that when visualization is useful, it is because visualization has a better trade-off than simply reading the data, simply using statistics alone, or simply relying on algorithms alone.
The ways to achieve a better trade-off include: (i) visualization may lose some information to reduce the human cost in observing and analyzing the data (e.g., overviews, glyphs, or other low-resolution visual representation), (ii) it may lose some information since the viewers have adequate knowledge to recover such information or can acquire such knowledge at a lower cost (e.g., deformed maps), (iii) it may preserve some information because it reduces the reconstruction distortion in the current and/or succeeding processes (e.g., external memorization), and (iv) it may preserve some information because the viewers do not have adequate knowledge to reconstruct such information or it would cost too much to acquire such knowledge.

% ----------
\begin{figure*}[t]
  \centering
  \includegraphics[width=178mm]{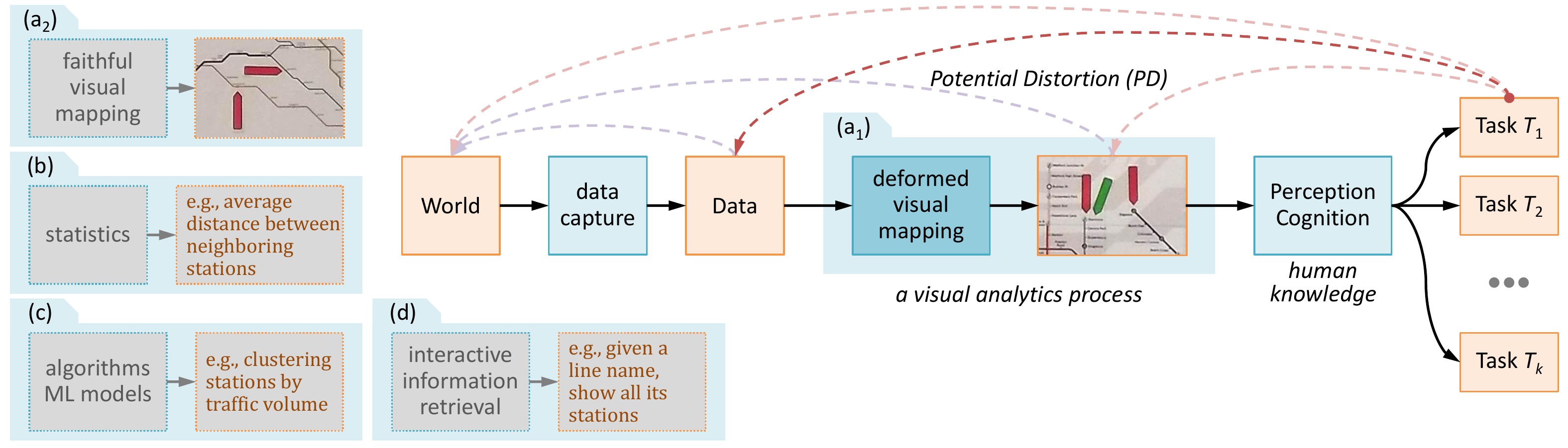}
  \caption{A visual analytics workflow features a general trend of alphabet compression from left (World) to right (Tasks). The potential distortion compares at an information space reconstructed based on the output with the original input information space. When we place different processes (i.e., (a$_1$), (a$_2$), (b), (c), and (d)), in the workflow, we can appreciate that statistics, algorithms, visualization, and interaction have different levels of alphabet compression, potential distortion, and cost. The figure is from Appendix B of \cite{Chen:2021:arXiv:E}.}
  \label{fig:Concept}
\end{figure*}

% ====================
\section{It can Explain Task- and User-Dependency}
\label{sec:TasksUsers}
Whilst hardly anyone in the visualization community would support any practice intended to deceive viewers, there have been many visualization techniques that inherently cause distortion to the original data.
The deformed London underground map is such an example.
The distortion in the commonly-used London underground map is largely caused by many-to-one mappings.
A group of lines that would be shown in different lengths in a faithful map are now shown with the same length.
Another group of lines that would be shown with different geometric shapes are now shown as the same straight line.
In terms of information theory, when the faithful map is transformed to the deformed, a good portion of information has been lost because of these many-to-one mappings.

In fact, there are many other forms of information loss. For example, when a high-resolution data variable (e.g., an integer in the range [0, 10,000]) is visually encoded as a bar in a bar chart that is restricted to a height of 1,000 pixels, about every 10 values are mapped onto the same height in terms of pixels.
It is unlikely that humans can precisely identify the height of each bar at the pixel resolution.
Likely a viewer may perceive a height of 833 pixels to be the same as one with 832 pixels or 834 pixels, which is also a many-to-one mapping.
When multivariate data records are encoded as glyphs, there is usually a significant amount of information loss. 
As we will discuss later in this paper, in volume visualization, when a sequence of $n$ voxel values are transformed to a single pixel value, as long as $n$ is a reasonably large value, a huge amount of information loss is almost guaranteed to happen.

Despite the ubiquitous phenomenon of information loss in visualization, it has been difficult for many of us to contemplate the idea that information loss may be a good thing. There are theories and guidelines in the field of visualization arguing for graphical integrity to prevent such information loss.
When one comes across an effective visualization but featuring noticeable information loss, the typical answer is that it is task-dependent, and the lost information is not useful to the task concerned.
When a visualization is evaluated, a common critique is about information loss, such as inadequate resolution, view obstruction, distorted representation, which are also characteristics of the aforementioned glyphs, volume rendering, and deformed metro maps respectively.

The common phrase that ``the appropriateness of information loss depends on tasks'' is not an invalid explanation. But on its own, this explanation is not adequate, because:
\begin{itemize}
    \item The appropriateness depends on many attributes of a task, such as the selection of variables in the data and their encoded visual resolution required to complete a task satisfactorily, and the time allowed to complete a task;
    \item The appropriateness depends also on other factors in a visualization process, such as the original data resolution, the viewer's familiarity of the data, the extra information that is not in the data but the viewer knows, and the available visualization resources;
    \item The phrase creates a gray area as to whether information loss is allowed or not, and when or where one could violate some principles such as those principles in \cite{Kindlmann:2014:TVCG}. 
\end{itemize}

Partly inspired by the above puzzling dilemma in visualization, and partly by a similar conundrum in economics ``what is the most appropriate resolution of time series for an economist'', Chen and Golan proposed an information-theoretic cost-benefit ratio for measuring various factors involved in visualization processes \cite{Chen:2016:TVCG}.
Because this cost-benefit ratio can measure some abstract characteristics of ``data'', ``visualization'', ``information loss'', ``knowledge'', and ``task'' using the most fundamental information-theoretic unit \emph{bit}, it provides a means to define their relationship coherently.
%

% Chen and Golan noticed that not only visualization processes lose information, but other data intelligence processes also lose information. For example, when statistics is used to down-sample a time series, or to compute its statistical properties, there is a substantial amount of information loss; when an algorithm groups data points into clusters or sort them according to a key variable, there is information loss; and when a computer system asks a user to confirm an action, there is information loss in the computational process \cite{Chen:2018:arXiv}. They also noticed that almost all decision tasks, the number of decision options is usually rather small. In terms of information theoretic quantities, the amount of information associated with a decision task is usually much lower than the amount of information associated with the data entering a data intelligence workflow.
% They concluded that this general trend of information reduction must be a positive thing for any data intelligence workflows.
% They referred to the amount of information reduction as \emph{Alphabet Compression} (AC) and made it a positive contribution to the \emph{benefit} term in Eq.\,\ref{eq:CBM-1a}.

Figure \ref{fig:Concept} shows an example of a simple visual analytics workflow, where at the moment, the visual analytics process is simply a visualization process, (a$_1$), for viewing a deformed London underground map. There can be many possible visualization tasks, such as counting the number of stops between two stations, searching for a suitable interchange station, and so on. From the workflow in Figure \ref{fig:Concept}, one can easily observe that the amount of information contained in the world around the entire London underground system must be much more than the information contained in the digital data describing the system.
The latter is much more than the information depicted in the deformed map.
By the time when the workflow reaches a task, the number of decision options is usually limited.
For example, counting the number stops may have optional values between 0 and 50.
The amount of information contained in the counting result is much smaller than that in the deformed map.
This evidences the general trend observed in \cite{Chen:2016:TVCG}.

One can simply imagine replacing the block (a$_1$) in Figure \ref{fig:Concept} with one of the other four blocks on the left, (a$_2$) for faithful visual mapping, (b) for statistics, (c) for algorithms, and (d) for interactive information retrieval.
This exercise allows us to compare the relative merits among the four major components of visual analytics, i.e., statistics, algorithms, visualization, and interaction \cite{Chen:2011:C}.

For example, statistics may be able to deliver a set of indicators about the London underground map to a user. In comparison with the deformed map, these statistical indicators contain much less information than the map, offering more AC contribution.
Meanwhile, if a user is asked to imagine how the London underground system looks like, having these statistical indicators will not be very helpful.
Hence statistics may cause more PD.

Of course, whether to use statistics or visualization may be task-dependent.
Mathematically, this is largely determined by both the PD and \emph{Cost} associated with the perception and cognition process in Figure \ref{fig:Concept}.
If a user tries to answer a statistical question using the visualization, it is likely to cost more than using statistics directly, provided that the statistical answer has already been computed or statistical calculation can be performed easily and quickly.

Whether to use statistics or visualization may also be user-dependent.
Consider a user \textbf{A} has a fair amount of prior knowledge about the London underground system, but another user \textbf{B} has little.
If both are presented with some statistics about the system (e.g., the total number of stations of each line), \textbf{A} can redraw the deformed map more accurately than \textbf{B} and more accurately than without the statistics, even though the statistical information is not meant to support the users' this task.
Hence to \textbf{A}, having a deformed map to help appreciate the statistics may not be necessary, while to \textbf{B}, viewing both statistics and the deformed map may help reduced the PD but may also incur more cost in terms of effort.
Hence visualization is more useful to \textbf{B}.

% This example echos the scenario presented in Figure \ref{fig:LondonMaps}, where we asked two questions:  Can information theory explain this phenomenon? Can we quantitatively measure some factors in this visualization process?
% If prior knowledge can explain the trade-off among AC, PD, and \emph{Cost} in comparing statistics and deformed map.
We can also extrapolate this reasoning to analyze the trade-off in comparing viewing the deformed map (more AC) and viewing the faithful map (less AC). 
Perhaps we can now be more confident to say that information theory can explain such a phenomenon.

To some readers, it may still be counter-intuitive to consider that information loss has a positive side. This is largely because the fact ``too much information loss will cause erroneous decisions'' is over-generalized to an incorrect perception ``information loss is not desirable.''
Recognizing the positive aspect of information loss is essential for asserting why visualization is useful as well as asserting the usefulness of statistics, algorithms, and interaction since they all usually cause information loss \cite{Chen:2019:CGF}.

% =================================================
\section{Information-Theoretic Formula of the Measure}
\label{app:InfoTheory}

In this section, we provide a concise summary of the mathematical definitions related to the cost-benefit measure proposed by Chen and Golan \cite{Chen:2016:TVCG}. From these definitions, those readers who are knowledgeable about the fundamental concepts in information theory can quickly notice that the cost-benefit measure is composed of two commonly-used information-theoretic measures. For those readers who are new to information theory, these definitions provide a pointer to relevant part of an information theory textbook (e.g., \cite{Cover:2006:book}).  
In addition, the original paper by Chen and Golan \cite{Chen:2016:TVCG} provides the mathematical rationale for the cost-benefit measure, while a recent book chapter by Viola et al. \cite{Viola:2019:book} provides the concept of ``visual abstraction'' with a mathematical explanation based on the cost-benefit measure. 

Let $\mathbb{Z} = \{ z_1, z_2, \ldots, z_n \}$ be an alphabet and $z_i$ be one of its letters.
$\mathbb{Z}$ is associated with a probability distribution or probability mass function (PMF) $P(\mathbb{Z}) = \{ p_1, p_2, \ldots, p_n \}$ such that
$p_i = p(z_i) \ge 0$ and $\sum_{1}^n p_i = 1$.
The \textbf{Shannon Entropy} of $\mathbb{Z}$ is:
\[
  \mathcal{H}(P(\mathbb{Z})) = - \sum_{i=1}^n p_i \log_2 p_i \quad \text{(unit: bit)}
\]

Here we use base 2 logarithm as the unit of bit is more intuitive in the context of computer science and data science. In a context that $\mathbb{Z}$ is unambiguously associated with $P$, one often write $\mathcal{H}(P(\mathbb{Z}))$ as $\mathcal{H}(\mathbb{Z})$ or $\mathcal{H}(P)$. In the literature of information theory, it is mostly written as $\mathcal{H}(P)$. However, for practical applications, writing it as $\mathcal{H}(P(\mathbb{Z}))$ or $\mathcal{H}(\mathbb{Z})$ helps remind us about the semantics of the alphabet $\mathbb{Z}$.

An alphabet $\mathbb{Z}$ may have different PMFs in different conditions.
Let $P$ and $Q$ be such PMFs. The \textbf{Kullback-Leibler divergence} (KL-Divergence), $\mathcal{D}_{KL}(P(\mathbb{Z})\|Q(\mathbb{Z}))$, measures the difference between the two PMFs in bits:
\[
  \mathcal{D}_{KL}(P(\mathbb{Z})\|Q(\mathbb{Z})) = \sum_{i=1}^n p_i \log_2 \frac{p_i}{q_i} \quad \text{(unit: bit)}
\]
$\mathcal{D}_{KL}(P||Q)$ is referred as the divergence of $P$ from $Q$.
This is not a metric since $\mathcal{D}_{KL}(P\|Q) \equiv \mathcal{D}_{KL}(Q||P)$ cannot be assured.

Consider a transformation $F: \mathbb{Z}_\text{in} \rightarrow \mathbb{Z}_\text{out}$, where $\mathbb{Z}_\text{in}$ is the input alphabet to $F$ with a PMF $P_\text{in}$ and $\mathbb{Z}_\text{out}$ is the output alphabet of $F$ with a PMF $P_\text{out}$ .
The term \emph{Alphabet Compression} (AC) in Eq.\,\ref{eq:CBR} is the difference between the input and output alphabet, $\mathcal{H}(\mathbb{Z}_\text{in}) - \mathcal{H}(\mathbb{Z}_\text{out})$.

Consider a reverse transformation $F^{-1}$ that attempts to reconstruct the input from the output. The reconstructed alphabet is expected to have a PMF different from that of the original input alphabet. We denote the reconstructed alphabet as $\mathbb{Z}'_\text{in}$ with a PMF $P_\text{in}$.
Thus the reverse transformation is $F^{-1}: \mathbb{Z}_\text{out} \rightarrow \mathbb{Z}'_\text{in}$.

Let the PMF of the original input alphabet be $Q(\mathbb{Z}_\text{in})$ and the PMF of the reconstructed alphabet be $P(\mathbb{Z}'_\text{in})$.
The term \emph{potential distortion} (PD) in Eq.\,\ref{eq:CBR} is defined using the KL-divergence as $\mathcal{D}_{KL}(P(\mathbb{Z}'_\text{in}) \| Q(\mathbb{Z}_\text{in}))$.

The mathematical definition of the qualitative formula in Eq.\,\ref{eq:CBR} is thus:
\begin{equation}\label{eq:CBR2}
    \frac{\textit{Benefit}}{\textit{Cost}} = \frac{\mathcal{H}(\mathbb{Z}_\text{in}) - \mathcal{H}(\mathbb{Z}_\text{out}) - \mathcal{D}_{KL}(\mathbb{Z}'_\text{in}||\mathbb{Z}_\text{in})}{\textit{Cost}}
\end{equation}
\noindent The fundamental measurement of the Cost is the energy required to perform $F$ and $F^{-1}$, while it can be approximated by a time or monetary measurement. Note that we use the simplified notation $\mathcal{H}(\mathbb{Z})$ in Eq.\,\ref{eq:CBR2} without explicitly mentioning a PMF $P$. 

As mentioned earlier, most measurement systems are not ground truth.
They are functions that map some reality to some quantitative values, in order to aid the
explanation of the reality and the computation of making predictions.
The cost-benefit measure in Eq.\,\ref{eq:CBR2} is one of such functions.
While the cost-benefit measure successfully captures trade-offs qualitatively
in data intelligence workflows, the measured values could shoot up toward infinity easily, hindering the reconstruction of the reality from the measured values.

Recently, Chen an Sbert proposed to replace the KL-divergence in Eq.\,\ref{eq:CBR2} with a bounded divergence measure \cite{Chen:2021:arXiv:T}, and Chen et al described two empirical studies for collecting practical data and using the data to evaluate several candidate divergence measures \cite{Chen:2021:arXiv:E}.
One of the empirical studies used two London underground maps, one abstract and one geographically-faithful, as the stimuli.
The other study used stimuli of volume visualization, where rendered images typically feature a huge amount of information loss and their interpretation relies extensively viewers' knowledge.

% ----------
\begin{figure*}[t]
    \centering
    \includegraphics[width=160mm]{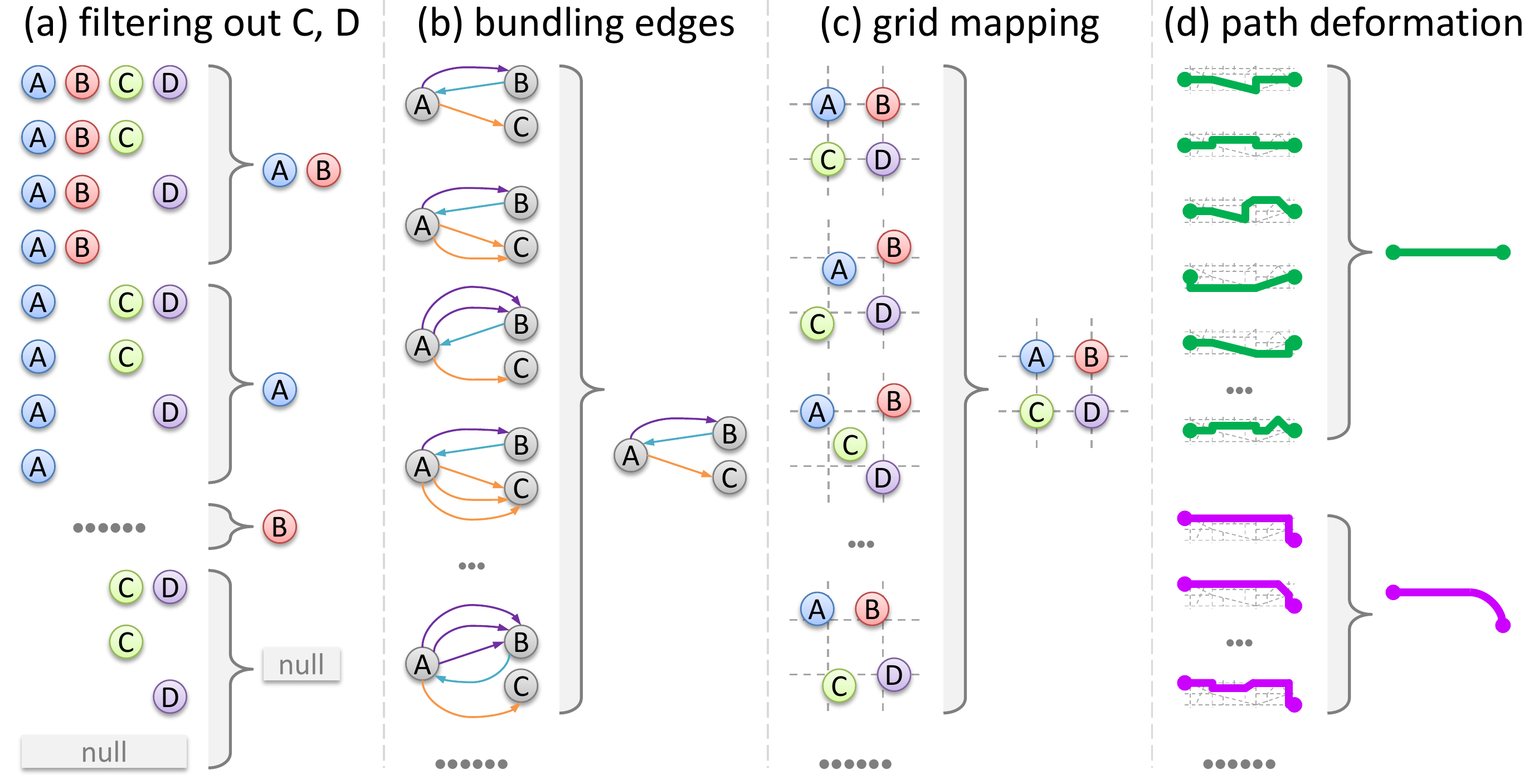}
    \caption{Four examples of entropy reduction or information loss in ODDV. (a) Whether a dataset may include any of four cities can be defined with an alphabet of 16 letters. When a filtering algorithm removes C and D from any input dataset, it creates a new alphabet with four letters, which has lower entropy. (b) The alphabet for encoding all possible connection patterns (up to $k$ edges) among three nodes contains many letters. Bundling edges with the same source and destination is a many-to-one mapping, which reduces entropy. (c) Grid mapping and path simplification, which are commonly-used design methods in ODDV, are also many-to-one mappings that cause information loss. The figure is from Appendix A of \cite{Tennekes:2021:CGF}.}
    \label{fig:Alphabet}
\end{figure*}

% ========================================
\section{Thinking in Alphabets: An Example}
\label{sec:ODDV}
Personally, it took me several years to accustom myself to thinking in an information-theoretic manner. Most of us are used to think about individual instances. Some can mentally reason with probability distributions, while some others can mentally reason with algebraic sets. Information theory asks us to think in both sets and probability distributions. Many of us may not feel ``intuitive'' or ``instinctive''' at the beginning.
However, as soon as one becomes accustomed to thinking in alphabets (including their PMFs), one feels liberated, a bit like the feeling when one first realizes being able to swim or ride a bike.
This section contains some text extracted from an appendix of a recent paper on origin-destination data visualization (ODDV) \cite{Tennekes:2021:CGF}, where we described some ODDV phenomena using the information-theoretic term of alphabet. 

In an ODDV process, before a viewer observes a visualization image, the viewer is uncertain about the OD dataset $D$ to be displayed.
In information theory, all mathematically-valid OD datasets form an alphabet $\mathbb{D}$, which is sometimes referred to as an information space.
A valid OD dataset is thus a letter of the alphabet, i.e., $D \in \mathbb{D}$.
Every letter in the alphabet is associated with a probability value, $p(D)$, indicating the likelihood that $D$ may appear. 

In a given context (e.g., rail commuting), many letters in $\mathbb{D}$ become impossible (e.g., about other mode of transport).
All possible datasets in this context constitute a sub-alphabet $\mathbb{D}_\text{ctx} \subset \mathbb{D}$.
In terms of Shannon entropy that measures the amount of uncertainty or information, the entropy of $\mathbb{D}_\text{ctx}$ is usually much lower than that $\mathbb{D}$.
Knowing the context enables a viewer to think, often unconsciously, using the  probability distribution for $\mathbb{D}_\text{ctx}$ instead that for $\mathbb{D}$.

When an algorithm is used to manipulate OD datasets in $\mathbb{D}_\text{ctx}$, it may further reduce the variations in $\mathbb{D}_\text{ctx}$.
For instance, as illustrated in Figure \ref{fig:Alphabet}, node filtering removes the possible variations of those nodes that are deleted if they occur in the data, while edge bundling creates a new alphabet that has fewer letters and thus fewer variations.
Grid-mapping and path simplification encode different geometrical variations using the same abstract representation.

In a given context, when a transformation $F$ is applied to all datasets in $\mathbb{D}_\text{ctx}$, it results in a new sub-alphabet $\mathbb{D}'_\text{ctx}$.
If $F$ features operations such as filtering, grouping, or distortion-based abstraction, $\mathbb{D}'_\text{ctx}$ will have less entropy than $\mathbb{D}_\text{ctx}$.
Entropy reduction implies information loss.
The usefulness of many visual designs in visualization, such as metro maps and many ODDV designs, evidence that information loss can have a positive impact, while challenging the traditional wisdom that a visual design needs to preserve all information in the data.
Sometimes one may argue that a visual design needs to preserve all information useful to a task. While the statement itself captures the task-dependent nature of visualization (but not the user-dependency), it is not ideal as it seems to imply a circular argument: ``\emph{a useful visual design shows useful information,}'' while neither usefulness can easily be defined.

% ----------
\begin{figure*}[t]
    \centering
    \includegraphics[width=160mm]{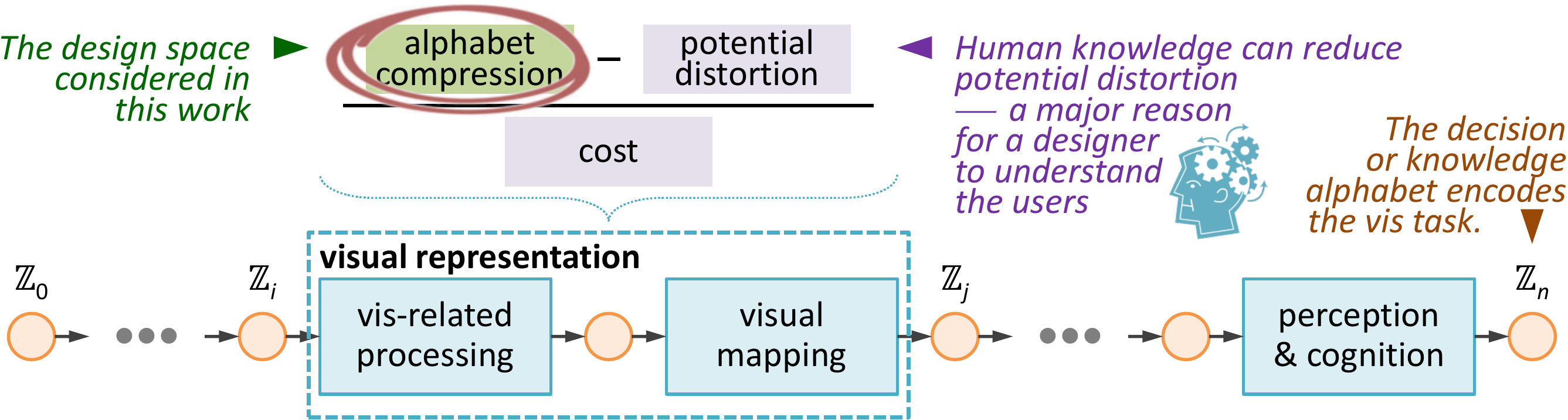}\\[2mm]
    \caption{A design space may categorize different options based on the amount of alphabet compression (i.e., losing information) and ways to achieve it. Too little information loss could increase the cost of the process and slowdown the progress towards the task objective. Too much information loss could increase potential distortion. Users' knowledge can alleviate potential distortion. The figure is from Appendix A of \cite{Tennekes:2021:CGF}. The phrase ``this work'' in the figure means \cite{Tennekes:2021:CGF}.}
    \label{fig:IT-DesignSpace}
\end{figure*}
% ----------

On the other hand, the cost-benefit analysis proposed by Chen and Golan has offered a mathematical explanation that such visual designs are cost-beneficial. 
According to the information-theoretic cost-benefit analysis \cite{Chen:2016:TVCG}, such information loss is part of the general trend of entropy reduction in a workflow from a data alphabet to a decision alphabet. Statistics, algorithms, visualization, and interaction in such a workflow all contribute to the entropy reduction (i.e., \emph{Alphabet Compression}). Hence entropy reduction itself is a merit rather than a demerit.
Without entropy reduction, there would be no decision.

In addition, entropy reduction at one stage helps reduce the \emph{Cost} of the stage or the succeeding stages.

Meanwhile, information loss may have a side-effect.
When a viewer observes an ODDV image that features filtering, grouping, distortion, or other data transformations that cause information loss, there is a possibility of misinterpretation (i.e., \emph{Potential Distortion}).

Using Figure \ref{fig:Alphabet}(d) as an example, a viewer who has little knowledge about metro maps, may interpret the path between the two stations is straight; a viewer, who understands concept of abstraction but knows little about the geography about that region, may make a random guess that the path can be of an arbitrary shape; or a viewer who lives nearby, may choose a shape that close to the reality.
Hence, the misinterpretation is viewer-dependent or user-dependent as we often say in visualization.

In many applications, some types of misinterpretations may not have a negative impact on the succeeding processes, where the transformations would converge to the same decisions regardless the variations of such interpretations.
As succeeding processes include tasks, this indicates that visualization is task-dependent. Hence, instead of stating that a piece of information is not useful to a task, information theory offers a mathematical definition of the usefulness, that is, (i) whether or not the extra information will lead to a different PMF $P_\text{with}$ of a decision alphabet from the PMF without the extra information $P_\text{without}$; and (ii) if $P_\text{with}$ and $P_\text{without}$ are different and if the ground truth PMF $Q$ is known, how $P_\text{with}$ and $P_\text{without}$ diverge from $Q$.

Once we appreciate that ODDV should enable entropy reduction and cannot avoid information loss unless the dataset is trivially simple, the question is then about \textbf{what} information to lose and \textbf{how} to lose information.
The principle design criteria are to reduce the potential distortions by maximizing the use of viewers' knowledge, reduce the costs of other human- and machine-processes that handle the data following the information loss, and reduce the negative impact on such processes.
In their paper \cite{Tennekes:2021:CGF}, Tennekes and Chen outline a design space categorized based primarily on the notions of \textbf{what} and \textbf{how}.
It focuses on different ways of alphabet compression as highlighted in Figure \ref{fig:IT-DesignSpace}, which also show that  the commonly-adopted wisdom of ``knowing the users and tasks'' is also supported by the information-theoretic reasoning.

% ====================
\section{Other Work and Future Work}
\label{sec:Other}
Making theoretical advancement is usually a long journey \cite{Chen:2017:CGA}. It requires the collective effort by an open-minded scientific community. An ideal theory has three main functions: (a) being able to explain phenomena in practice, (b) being able to offer interpretable measurement, and (c) being able to make dependable prediction. We do not have an ideal theory for visualization and visual analytics yet, nor should we use these criteria to block the progression of theoretical advancement in the field. Not many people can produce a wonderful theory in one publication. My own experience indicates a rather slow and iterative process of observation, reading, understanding, formulation, self-doubting, collaboration, writing, and improvement. Table \ref{tab:Work} lists some of the theoretical research by me and my colleagues. They are small steps made in each case toward a long-term aim of having some ideal theories for visualization and visual analytics.

The cost-benefit ratio proposed by Chen and Golan is relatively successful in explaining phenomena in visualization and visual analytics, and potentially in some other fields including perception and cognition, language development, news media, and machine learning \cite{Chen:2020:OUP}. We should continue to look for phenomena that can as well as cannot be explained by the cost-benefit analysis. To confirm a theoretical postulation usually needs numerous pieces of evidence unless there are other confirmed theories that can be used to prove the postulation. To falsify a theoretical postulation usually needs only one piece of solid evidence. Discovering such evidence usually paves the way for a new theoretical postulation.

As mentioned earlier, the cost-benefit ratio may not be intuitively interpretable due to the unbounded PD component \cite{Chen:2021:arXiv:T,Chen:2021:arXiv:E}. Since a measuring function is a form of abstraction, using Chen and Golan's own terms, the potential distortion and cost of using the current mathematical definition may be rather high, and we should seek to provide a better measurement function. Many measurement systems in the history underwent improvement over years and decades, such as temperature scales and seismological scales.

There is some limited progress towards making prediction, mainly using the qualitative version of the cost-benefit ratio \cite{Chen:2019:CGF}. Hopefully, there will soon be some concrete advancement in the aspect of measurement, which will facilitate more concrete methodological advancement for making quantitative prediction.

\begin{table*}[hbt]
\caption{Some publications related to the information-theoretical cost-benefit measure.}
\centering
\begin{adjustbox}{max width=\textwidth}
\begin{tabular}{@{}lll@{}}
\toprule
%\multicolumn{2}{c}{Name} \\
% \cmidrule(r){1-2}
\textbf{Citation} & \textbf{Contribution} & \hspace{3cm}\textbf{Brief Description} \\
\midrule
Chen \& J\"{a}nicke \cite{Chen:2010:TVCG} & \emph{Explanation} &
    A few phenomena (e.g., overview first, redundancy, motion parallax) \\
& \emph{Measurement} & Three measures (i.e., VMR, ILR, DSU)\\
& \emph{Prediction} & A few information-theoretical laws\\
& \emph{Observation} & Interactive visualization vs. data processing inequality; visualization vs. compression\\ 
\midrule
Chen et al. \cite{Chen:2014:CGF} & \emph{Explanation} &
    Explaining multiplexing phenomena in visualization based on information space\\
& \emph{Observation} & Categorization of multiplexing phenomena in visualization\\
\midrule
Chen \& Golan \cite{Chen:2016:TVCG} & \emph{Explanation} &
    Trade-off phenomena in data analysis and data visualization\\
& \emph{Measurement} & The information-theoretic formula of the cost-benefit ratio\\
& \emph{Observation} & Categorization of visualization tasks based on space complexity\\
\midrule
Tam et al. \cite{Tam:2017:TVCG} & \emph{Measurement} &
    Estimating human knowledge used in visualization-assisted machine learning\\
\midrule
Kijmongkolchai et al. \cite{Kijmongkolchai:2017:CGF} & \emph{Measurement} &
     Estimating the benefit and cost of visualization processes using an empirical study\\
\midrule
Chen \cite{Chen:2018:arXiv} & \emph{Explanation} &
    The important role of interaction for humans to use their knowledge in data intelligence\\
& \emph{Measurement} & Methods for estimating human knowledge by observing interactions\\
\midrule
Chen \& Ebert \cite{Chen:2019:CGF} & \emph{Predication} &
    A qualitative methodology for improving visual analytics workflows in a systematic manner\\
\midrule
Chen et al. \cite{Chen:2019:TVCG} & \emph{Explanation} &
    Analyzing the cost-benefit of visualization processes in different virtual environments\\
& \emph{Prediction} & Answering questions from a workshop and a few predictions in its appendices\\
\midrule
Chen \cite{Chen:2020:OUP} & \emph{Explanation} &
    Trade-off phenomena in machine learning, perception/cognition, language development, etc.\\
& & Also including a definition of ``Data Science'' differing from ML-focused definitions\\
\midrule
Viola et al. \cite{Viola:2019:book} & \emph{Explanation} &
    A mathematical explanation of the concept and phenomena of ``Visual Abstraction''\\
\midrule
Streed et al. \cite{Streeb:2021:TVCG} & \emph{Explanation} &
    Comparing the explainability of many arguments or theoretical postulations in the literature\\
\midrule
Tennekes \& Chen \cite{Tennekes:2021:CGF} & \emph{Explanation} &
    Information loss in origin-destination data visualization\\
& \emph{Prediction} & A design space for searching ``predictions'' based on categorization of information loss\\
\midrule
Chen \& Sbert \cite{Chen:2021:arXiv:T} & \emph{Measurement} &
    A theoretical analysis of candidate divergence measures for the cost-benefit ratio\\
\midrule
Chen et al. \cite{Chen:2021:arXiv:E} & \emph{Measurement} &
    A data-driven analysis of candidate divergence measures for the cost-benefit ratio\\
\bottomrule
\end{tabular}
\end{adjustbox}
\label{tab:Work}
\end{table*}

% ====================
% \section*{Acknowledgments} % The \section*{} command stops section numbering
%
% Text

%\addcontentsline{toc}{section}{Acknowledgments} % Adds this section to the table of contents

%----------------------------------------------------------------------------------------
%	REFERENCE LIST
%----------------------------------------------------------------------------------------
\phantomsection
\bibliographystyle{unsrt}
\bibliography{references}

%----------------------------------------------------------------------------------------

\end{document}